\documentclass[preprint,showpacs,preprintnumbers,amsmath,amssymb]{revtex4}
\usepackage{graphicx}% Include figure files
\usepackage{dcolumn}% Align table columns on decimal point
\usepackage{bm}% bold math

\begin{document}

\title{Secondary nuclear fragment beams for investigations of relativistic fragmentation of light radioactive nuclei using nuclear photoemulsion at Nuclotron}

\author{P.A. Rukoyatkin}
   \email{rukoyat@sunhe.jinr.ru}
   \affiliation{Joint Insitute for Nuclear Research, 141980 Dubna, Russia} 
\author{L.N. Komolov}
   \affiliation{Joint Insitute for Nuclear Research, 141980 Dubna, Russia} 
\author{R.I. Kukushkina}
   \affiliation{Joint Insitute for Nuclear Research, 141980 Dubna, Russia} 
\author{V.N. Ramzhin}
   \affiliation{Joint Insitute for Nuclear Research, 141980 Dubna, Russia} 
\author{P.I. Zarubin}
   \affiliation{Joint Insitute for Nuclear Research, 141980 Dubna, Russia}

\begin{abstract}
\indent  Slowly extracted relativistic beams of light nuclei and a beam transportation line net system constitute a good base for secondary nuclear beams forming at the LHE accelerator facility. A recent years activity in the field at the Laboratory is connected with a project on study light nuclei structure by means the emulsion technique \cite{Bradnova,Adamovich1}. The paper shortly summarizes results of the work.\par
\indent \\
\indent \par
\indent DOI: 10.1140/epjst/e2008-00802-0\par
\end{abstract}
  %    {PACS-key}{21.45.+v} \and
   %   {PACS-key}{23.60+e} \and
    %  {PACS-key}{25.10.+s}  

\maketitle
\section{INTRODUCTION}

\indent A superconducting accelerator of nuclei, the Nuclotron, is now a basic facility of the Laboratory of High Energies (LHE) \cite{Smirnov}. The machine is intended to carry out fundamental investigations in the relativistic nuclear physics area at energies up to 6A~GeV as well as to solve various applied tasks. An existing injection subsystem provides the facility with a wide set of nuclei up to Fe. Accelerated beams can be immediately used for internal target based experiments and are slowly extracted out of the machine and transported through a broad experimental area to external physical setups \cite{Issinsky}. The superconducting realization of the ring makes possible to realize practically a continuous extracted beam. A ten seconds extraction duration has been recently demonstrated \cite{Volkov}.\par

\indent Relativistic beams of light nuclei slowly extracted from Nuclotron and a developed beam line system of the facility constitute a ground for secondary nuclear fragment beams forming in-flight. The possibility was widely practiced at the LHE to create some specific secondary beams ($^3$H, $\mathbf{n}$, $n$, $\mathbf{p}$). A short summary with references on the subject is given in \cite{Rukoyatkin}.\par

\indent A resent years activity in the field was initiated by the mentioned physical program of emulsion irradiations in beams of relativistic nuclei named the BECQUEREL project. It is destined to study in detail the processes of relativistic fragmentation of light radioactive and stable nuclei. The expected results would make it possible to answer some topical questions concerning the cluster structure of light nuclei. The Laboratory has accumulated considerable experience in the nuclear emulsion method supported it over a period of many years. In spite of seeming 'obsoleteness' an emulsion stack owning to the best spatial resolution stays a modern low cost tabletop 4$\pi$-detector for a set of relativistic nuclear physics processes (f.e. \cite{Andreeva}). According to the program a series of emulsion exposures were carried out both in primary beams ($^{10,11}$B, $^{14}$N) and in secondary fragment beams enriched with $^3$H~$/$~$^6$He, $^{7,9}$Be, $^{9,10}$C, $^8$B and $^{12}$N nuclei. The exposures were carried out at kinetic energies 1$-$2A~GeV.\par 	

\indent Forming of the secondary beams is based on peripheral nucleus fragmentation reactions or charge-exchange reactions. In the first case secondary fragments produced are describe in the projectile frame by an isotropic normal momentum distribution with small r.m.s. $\sigma_1$ as against to their mass $M_1$ and in the laboratory frame r.m.s. of plane angle $\theta$ and of relative momentum $\sigma$ are
\begin{equation}
\begin{split}
\sigma_{\theta}\simeq\frac{\sigma_1}{A_1p_0},~~~~~~\sigma_{\delta}\simeq\frac{\sigma_1}{\beta_0M_1},
\end{split}
\end{equation}
where $\sigma_1=\sigma^*\sqrt{A_1(A_0-A_1)/(A_0-1)}$ \cite{Goldhaber}, $\sigma^*\simeq90$~~MeV/$c$, $A_0$, $A_1$ are projectile and fragment mass numbers, $p_0$, $\beta_0$ are projectile momentum per nucleon and velocity. As it follows from the second relation the lighter secondary fragments the larger their momentum spread (larger tails) what is probably just a reason of observable prevailing of helium isotopes admixture in heavier fragment beams. Numerically, f.e., $\sigma_{\delta}\simeq1.8\%$ for $^{10}$B$\rightarrow^8$B whereas for $^{10}$B$\rightarrow^3$He $\sigma_{\delta}\simeq5.4\%$.\par

\section{Using the beam transportation lines for nuclear fragment beams forming}

\begin{figure}
    \includegraphics[width=4in]{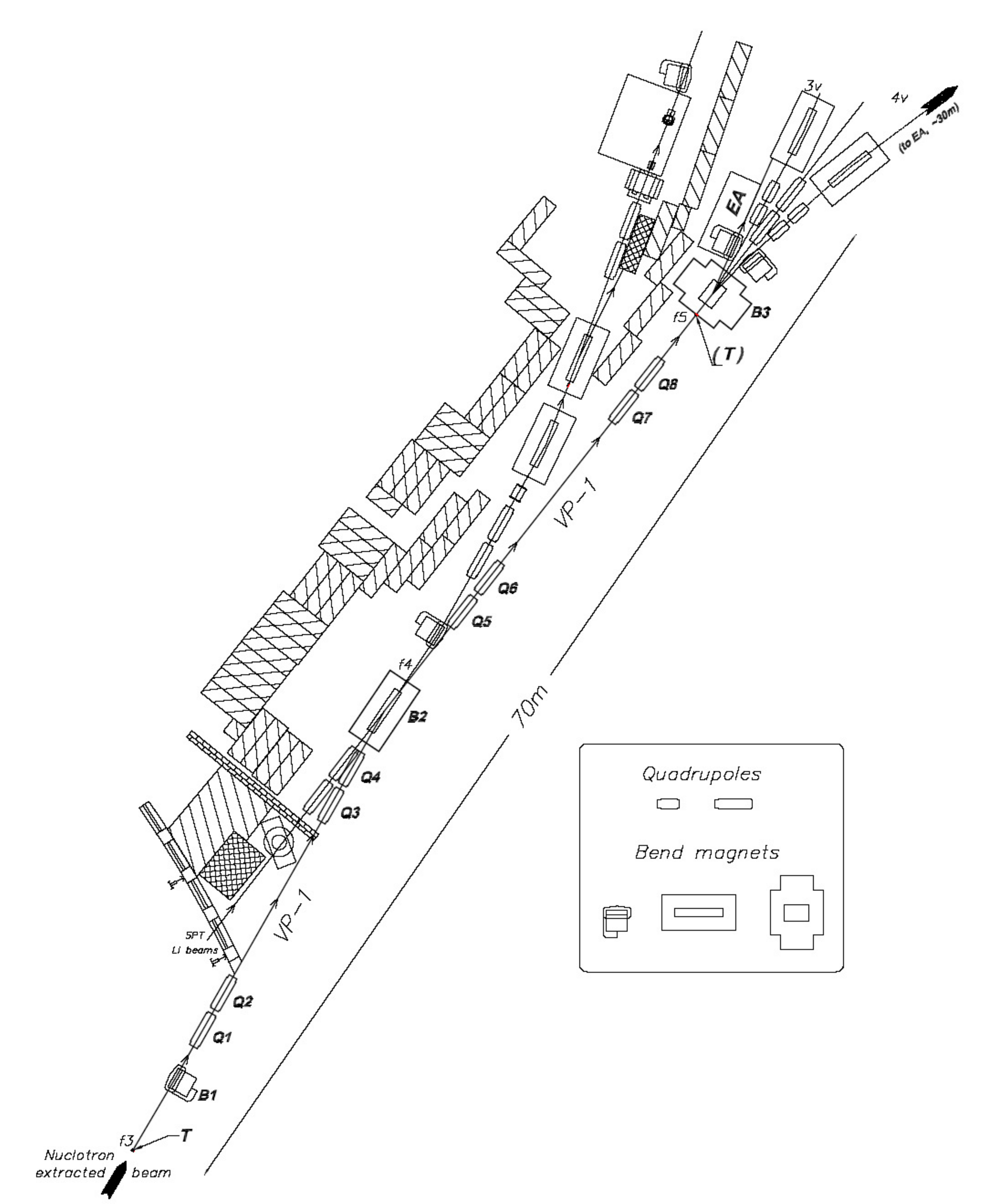}
    \caption{\label{Fig:1} Scheme of an experimental area at Nuclotron extracted beam. The main part of the VP-1 beam line and the 4v line were applied to form secondary nuclear fragment beams for emulsion irradiations. T$-$stripping target positions. EA$-$emulsion assemblage positions.}
    \end{figure}
	
\indent The VP-1 and 4v lines were applied for beam forming of secondary nuclear fragments for the emulsion program. The first experiments were carried out at the 4v beam line which is concerned below in section 4. The most of exposures have been done using the VP-1 one. The VP-1 beam line is the main channel of a fir-tree like beam arrangement net \cite{Issinsky} transporting accelerated extracted beams to external physical setups. It consists of 16 magnet elements and has the total length of $\simeq$100~m. A shield allows one to work on the VP-1 at any intensities available at the facility. The main part of the of the line is shown in Fig.~\ref{Fig:1}. Normally the VP-1 line transfers primary beams from the f3 point (the final point of the Nuclotron slow extraction system) through experimental halls to a beam dump. Three crossovers in both planes are formed at the f4, f5 (and f6, out of the figure) points. The horizontally bending magnets B1 and B2 operates, the B3 is switched off. The standard polarity of the Q1-Q4 and Q5-Q8 lenses quartets in the horizontal plane is FDDF. To use the line for secondary fragments beam forming a certain dispersiveness has to be obtained. Like a rough guess following from (1) a sufficient resolution value (r) can be taken as $r\simeq\sigma^*/mA_0$, where m is a nucleon mass. Thus for the case of boron/carbon beams fragmentation an $r\simeq1\%$ should be realized.\par

\indent As a beam line optics analysis showed a dispersion introduced by the B2 magnet can be retained at some level at the f5 point and successively yet increased at a corresponding incorporation of the B3. Optimization of the structure was carried out using a ratio $R(z)$ defined in the following way:
\begin{equation}
\begin{split}
R(z)=\frac{r_{16}(z)}{2\sigma_x(z)},
\end{split}
\end{equation}
where $z$ is the distance along beam line, $r_{16}$ is the linear dispersion and $\sigma_x$ is the r.m.s. size of the monochromatic beam in the dispersion plane.\par

\begin{figure}
    \includegraphics[width=5in]{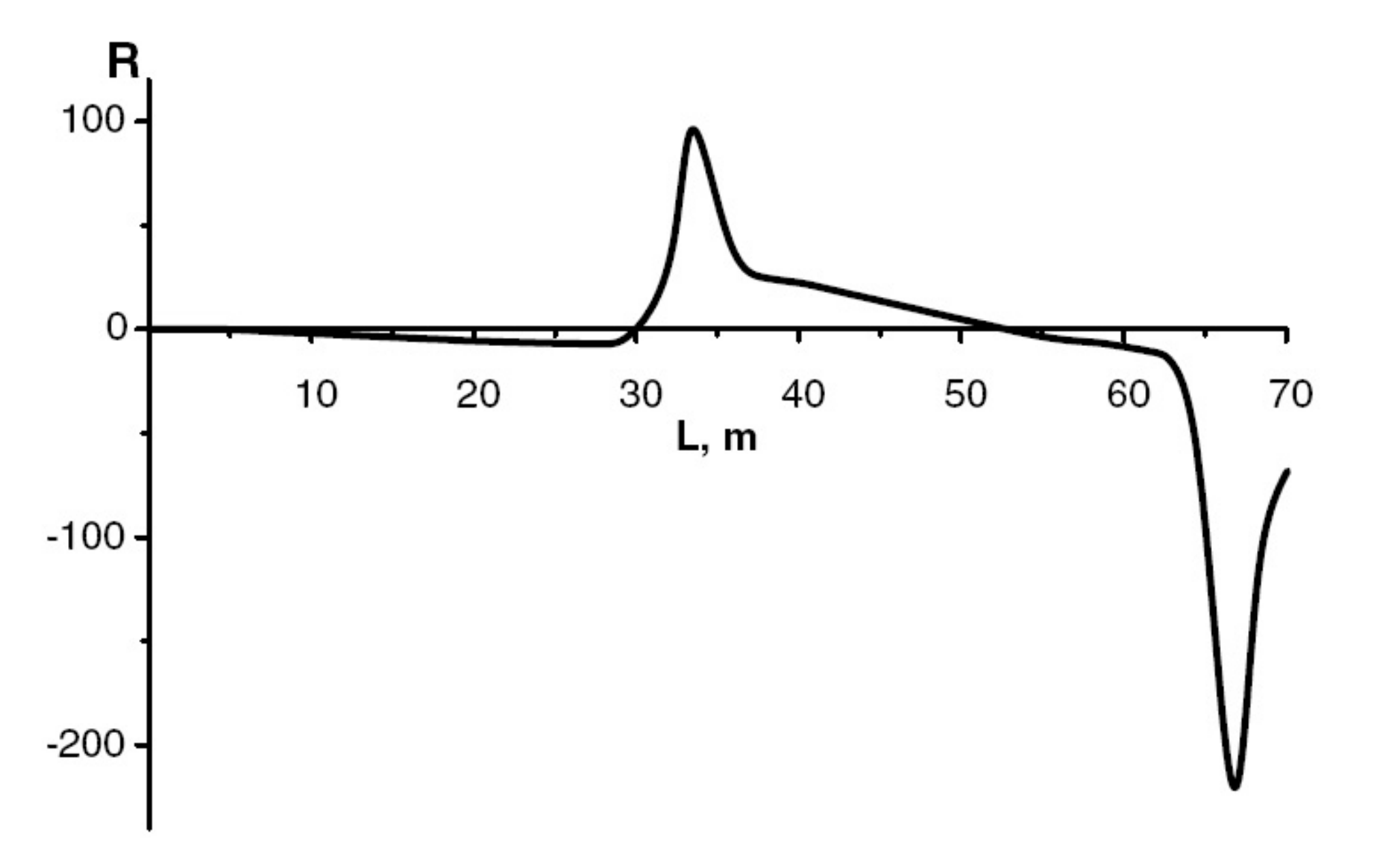}
    \caption{\label{Fig:2} A calculated beam line resolution function.}
    \end{figure}

\indent A solution of the optimization task was taken as an operating regime base. An example of an optimized $R(z)$ function is given in Fig.~\ref{Fig:2}. At the peak region the function has very strong dependence on its variables: f. e. varying the Q4-Q8 quartet gradients on a percent varies the $R(z)$ up to 30\%. Experimentally an $R(z)$ function values were estimated measuring shifts of primary beam positions at a small varying of the beam momentum. Such variation was realized by a degrader placed into a beam at the system entrance. In practice the absolute values of gradients can be set with a finite precision. That is why in a restricted time frame of a beam run a realized $R(z^*)$ value at a desired $z^*$ position was usually less in comparison with the
calculation. By beam run estimations working $R'$s values were 70$-$100.\par

\begin{figure}
    \includegraphics[width=5in]{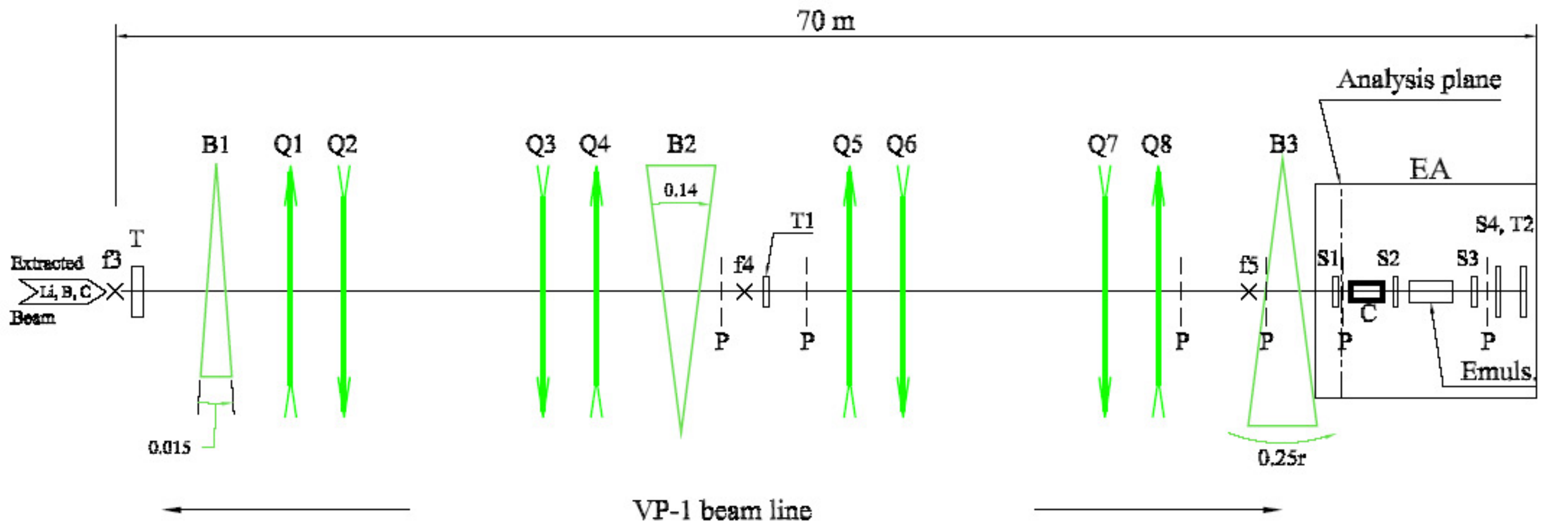}
    \caption{\label{Fig:3} Optical scheme of the beam line and detectors layout. Dashed marks (P) designate multiwire profilemeters. The S1 counter of a diagnostic detector assemblage (EA) is placed close to a maximum position (analysis plane) of the beam line resolving function $R(z)$.}
\end{figure}

\indent A relative disposition of the beam line components and diagnostic detectors is shown in Fig.~\ref{Fig:3}. Primary nuclei beams interacted with a target placed at the f3 point. At resent experiments projectiles were $^{6,7}$Li, $^{10}$B, $^{12}$C at momenta of 1.7$-$2.7A~GeV/$c$. Polyethylene targets of 4$-$8~cm in thickness were used.\par

\section{Methodical details}

\indent The main monitoring tool of primary nuclei beams was a set of profilemeters (P). Each profilemeter is a multiwire ionization chamber measuring beam shapes both in the horizontal and the vertical planes (30X$+$30Y channels).\par

\indent A detectors assemblage (EA) was placed just downstream a designed maximum position of the channel resolving function (analysis plane). It included a set of scintillation counters, multiwire profilemeters and a collimator. The collimator was used at some emulsion exposures. At beam testing the S1 counter of 10mm in width was used as a momentum slit tag. The S2 fitting an entrance section of an emulsion stack (100$\cdot$10~mm$^2$) was intended for emulsion load monitoring and also served as a beam analyzer. Typical loads were 0.5$-$2$\cdot$10$^5$ secondary nuclei per a stack. The S3 counter with a 8~mm thick plastic scintillator was used as the main dE/dx beam content analyzer. An additional triggering counters (S4) was placed 5~m downstream the analysis plane. At the last run a scintillation TOF pair (T1, T2) of was included into the assemblage. The pair had proper time resolution of $\simeq$0.24~ns and was spaced off $\simeq$37~m.\par

\indent An experimental spectrum with m peaks was approximated by a function of $4m+2$ varied parameters $\mathbf{p}=(\underbrace{...,S_i,c_i,w_i,k_i,...}_{m},\sigma,y_0)$
\begin{equation}
\begin{split}
f(x,\mathbf{p})=\frac{1}{\sqrt{2\pi}\sigma}\int\sum_{i=1}^m S_iv_i(x-t)e^{-t^2/2\sigma^2}dt+y_0,
\end{split}
\end{equation}
where $v_i(y)=v((y-c_i)/w_i-\Delta(k_i,\beta),k_i,\beta)$ is the Vavilov distribution function with parameters $k_i$ and $\beta$, $S_i$ is the square of a peak, $c_i$ is its maximum position, $w_i$ is width parameter.\par

\indent Each of the primary beams was conducted through the channel twice. At first a low intensity ($<10^5~p/s$) beam was passed for calibration of the analyzers. Then beam passed through the production target was conducted at working intensity $10^8-10^9$ nuclei per accelerator cycle. After that fields of the channel magnetic elements were strictly changed accordingly to a ratio of selected secondary rigidity and projectile one.\par

\indent The first stage of secondary beam selection was realized by the head part of the VP-1 channel. This part operated practically at a standard regime. At this stage the main contribution to dispersion originated from the B2 magnet with a bend angle $\simeq140~$mr. Primary selection of a momentum interval was determined by free aperture of the Q1$-$Q5 quadrupoles. Final dispersive parameters of the system were formed by the successive Q5$-$Q8 and B3 ($\phi\simeq250~$mr) elements accordingly a particular realization of the optimized base regime. After the first stage a momentum spread was $\sigma\simeq5\%$ as one followed from a simulation carried out by mean of the codes \cite{Carey}. After the whole channel with a 10~mm slit at it narrowed up to 0.5\%.\par

\section{Main results}

\begin{figure}
    \includegraphics[width=5in]{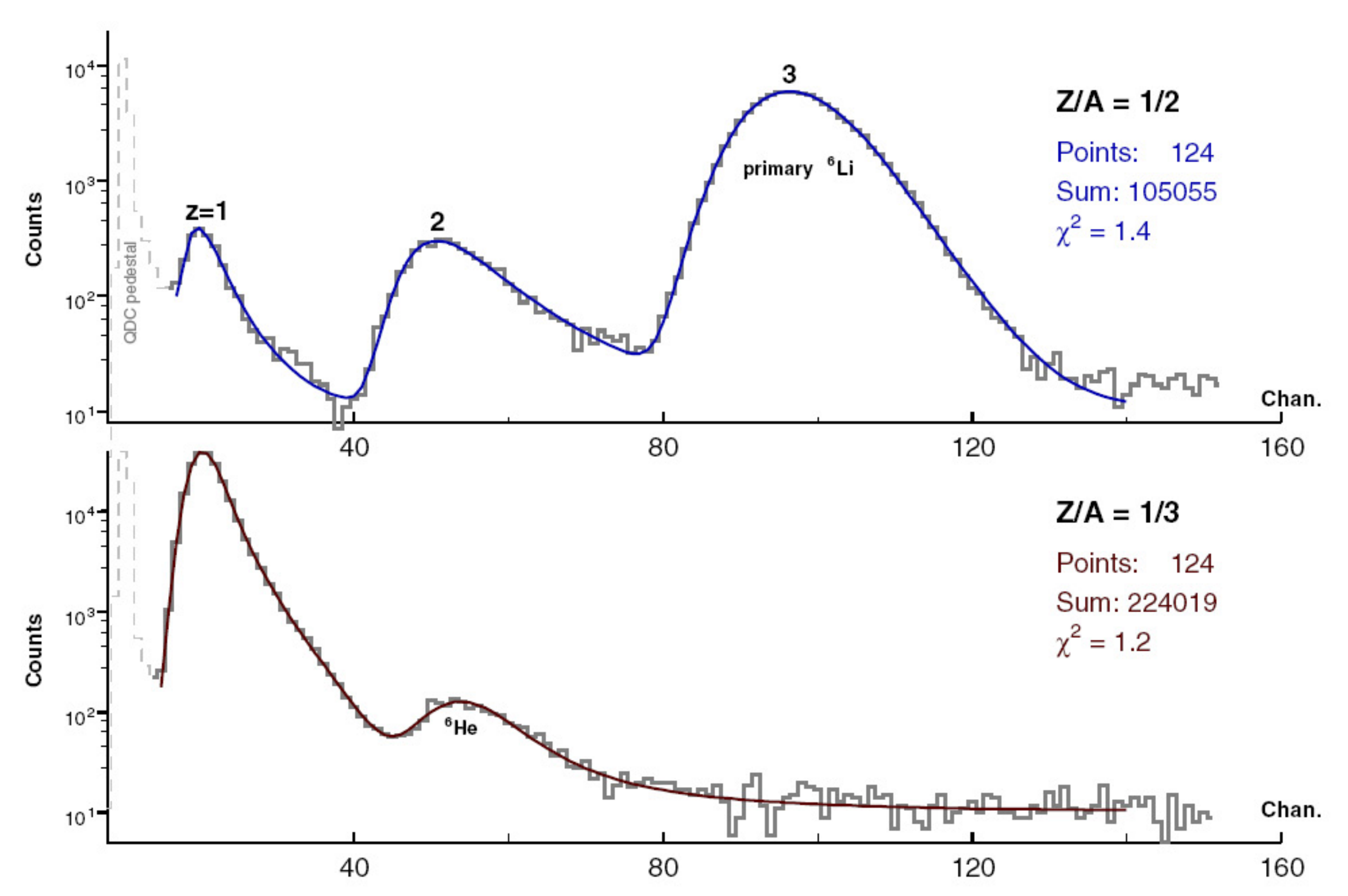}
    \caption{\label{Fig:4} Energy deposition spectra in a 8~mm scintillator registered in secondary beams produced out of 2.7A~GeV/$c$ $^6$Li nuclei interaction. Top: at beam line tune to rigidity of fragments with Z/A$=$1/2 (scale binding). The predominant peak $\mathbf{3}$ corresponds to the primary nuclei passed the target without interactions. Bottom: at Z/A$=$1/3. Smooth curves are fits of type (3).}
\end{figure}

\indent The first test run on forming a nuclear fragment beam for emulsion irradiation was carried out on a $^6$Li beam accelerated by the Synchrophasotron \cite{Adamovich2}. At the run the primary beam was conducted to a target at the f5 point and the 4v beam line was used (Fig.~\ref{Fig:1}), only head part is shown). The main parameters of the forming are summarized in Table 1. Initially a low intensity primary beam ($\simeq10^5$~s$^{-1}$) passing a target and partially fragmetizing was carried out through the beam line for calibration of a registering chain PMT$+$QDC of an 8~mm plastics analyzer (top plot in Fig.~\ref{Fig:4}). A similar procedure was executed at all succeeded experiments. After that fields of the beam line elements were strictly increased by factor 3/2 what corresponded to rigidity of $^3$H/$^6$He nuclei. The bottom plot in the figure represents a contents of the secondary beam. Observed relation of the $^6$He component presence to the tritium one $\simeq0.8\%$, total yield of the secondaries into the acceptance $\simeq2\cdot10^{-4}$ per an incident $^6$Li nucleus. Divergency of the beam was monitored by means of two MWPC. An on-line estimated value of the divergency in the plane perpendicular to emulsion layers was not exceed 2.5~mr what has been in agreement with one obtained on scanned emulsion tracks.\par

\begin{table}
\caption{\label{Tabel:1}Forming conditions of secondary beams at the 4v beam line on the 6Li nuclei base.}
\begin{tabular}{|c|c|c|c|c|c|c|}
\hline\noalign{\smallskip}
\multicolumn{3}{|c|}{Primary beam at target}&&\multicolumn{3}{c|}{4v beam line}\\
\noalign{\smallskip}\hline\noalign{\smallskip}
Momentum & A~GeV/$c$ & 2.7 & & Mag. elem. &  & 10 \\
Size, $\sigma_x/\sigma_y$ & mm & 3/9 & & Length & m & 51 \\
Diverg., $\sigma_{\theta_x}/\sigma_{\theta_y}$ & mr & 6/2 & & Accept., $\Delta\Sigma$ & $\mu$sr & $\simeq60$ \\
Intensity & nucl./spill & $3-5\cdot10^7$ & & ~~~~~~$\delta p$ & \% & $\pm2$ \\
Spill tretch & ms & $\simeq500$ & & Target (Plex.) & g/cm$^2$ & 5.1 \\
\noalign{\smallskip}\hline
\end{tabular}
\end{table}

\begin{figure}
    \includegraphics[width=4in]{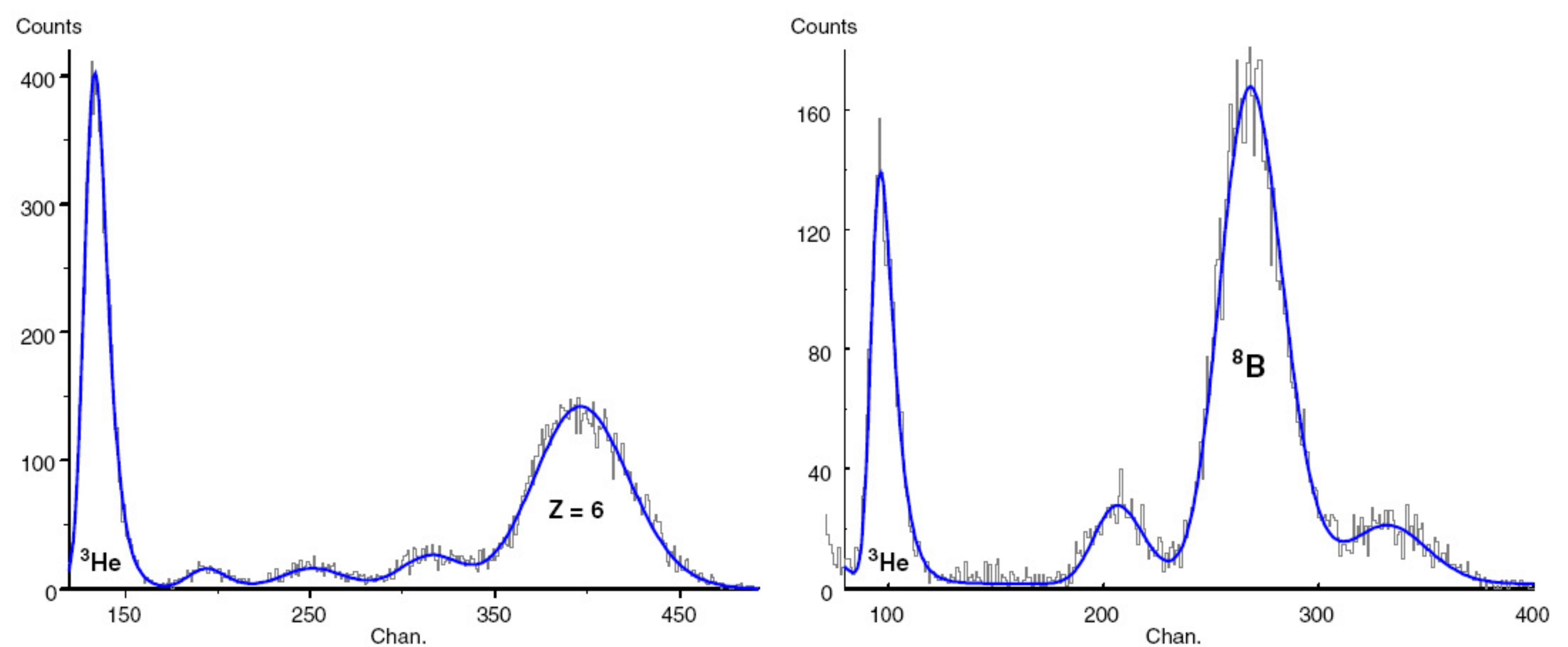}
    \caption{\label{Fig:5} Secondary beams produced by 2.0A~GeV/$c$ $^{12}$C fragmentation at a beam line tune to the $^9$C fragment rigidity (left) and by 2.0A~GeV/$c$ $^{10}$B at a tune to $^8$B (right).}
\end{figure}

\begin{figure}
    \includegraphics[width=4in]{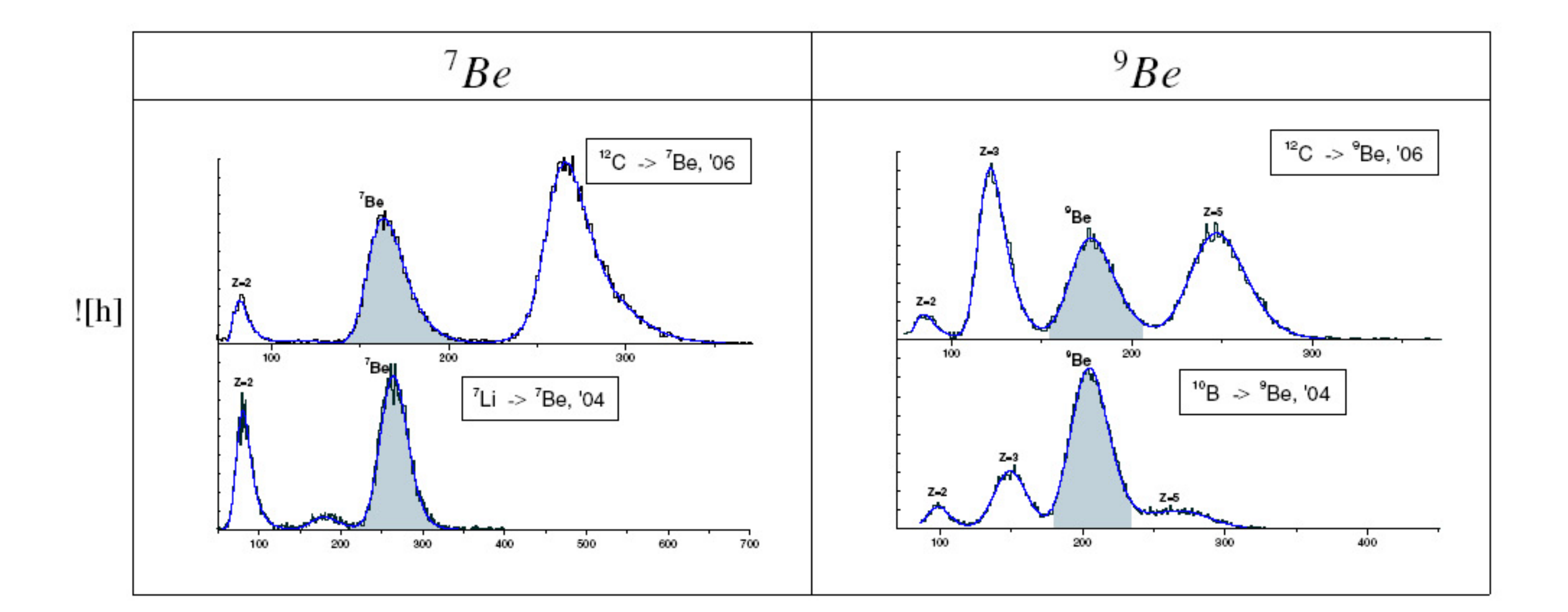}
    \caption{\label{Fig:6} Content comparison of secondary beams enriched with beryllium isotopes produced by different primary projectiles. Spectra were obtained at different runs $-$ scales (amplification) are not the same.}
\end{figure}

\indent Further works on forming of secondary nuclear beams for emulsion exposures were carried out using the VP-1 beam line as it has been described above. The exposures were aimed to take data on $^7$Be, $^9$Be, $^8$B, $^9$C, $^{12}$N nuclei. Each of corresponding secondary beams obtained was a natural blends of nuclei determined by a beam line resolution, adjacent fragments proximity on rigidity, their momentum spread (estimated by Eq. (1)) and fragmentation channels probabilites. Spectra representing content of the beams are given in Figs.~5$-$8. Table~2 summarizes numerical values relating to the spectra. In the cases of $^{7,9}$Be, $^8$B beams a desired fragment had the greatest contribution. Presence of the main Z$=$6 component at a $^9$C tune was some above 50\%. The component could include a certain part of the $^{10}$C admixture unrecognized by used detectors. A definite presence of a sattelite Z$=$6 component at a $^8$B tune indicates a practical feasiblity to obtain $^{10}$C from the $^{10}$B beam via a charge-exchange mechanism. Less clean experimental conditions have been realized at a tune to the charge-exchange mode $^{12}$C$+$A$\rightarrow^{12}$N$+...$. Nevertheless, a certain part of Z$=$7 events in the secondary beam is recognized (Fig.~\ref{Fig:8}). In a conclution, one should remark resecent papers \cite{Peresadko,Artemenkov,Stanoeva} with some physical results obtained on emulsions exposed in the formed beams.\par

\begin{table}
\caption{\label{Tabel:2}Secondary beams formed at the VP-1 beam line for emulsion exposures.}
\begin{tabular}{|c|c|c|c|c|c|c|c|c|c|c|}
\hline\noalign{\smallskip}
 & & p, & Frag. & & \multicolumn{6}{c|}{Relation of main components}\\
\cline{6-11}
 & Proj. & A GeV/$c$ & (tune to) & Fig. & Z$=$2 & 3 & 4 & 5 & 6 & 7 \\
\noalign{\smallskip}\hline\noalign{\smallskip}
1 & $^7$Li & 1.7 & $\mathbf{^7Be}$ & (6) & 0.4 & 0.1$<$ & $\mathbf{1}$ & & & \\
\noalign{\smallskip}\hline\noalign{\smallskip}
2 & $^{10}$B & 2.0 & $\mathbf{^9Be}$ & (6) & 0.1$<$ & 0.3 & $\mathbf{1}$ & 0.1 & & \\
\noalign{\smallskip}\hline\noalign{\smallskip}
3 & $^{10}$B & 2.0 & $\mathbf{^8B}$ & (5) & 0.3 & & 0.1 & $\mathbf{1}$ & 0.14 & \\
\noalign{\smallskip}\hline\noalign{\smallskip}
4 & $^{12}$C & 2.0 & $\mathbf{^9C}$ & (5) & 0.7 & \multicolumn{3}{c|}{$\sum\simeq0.2$} & $\mathbf{1}$ & \\
\noalign{\smallskip}\hline\noalign{\smallskip}
5$^*$ & $^{12}$C & 2.0 & $\mathbf{^{12}N}$ & (8) & & & & & & $\mathbf{0.08^{**}<}$ \\
\noalign{\smallskip}\hline\noalign{\smallskip}
6$^*$ & $^{12}$C & 2.0 & $\mathbf{^7Be}$ & (6) & 0.3 & & $\mathbf{1}$ & \multicolumn{2}{c|}{$\sum\simeq2$} &\\
\noalign{\smallskip}\hline\noalign{\smallskip}
7$^*$ & $^{12}$C & 2.0 & $\mathbf{^9Be}$ & (6) & 0.1 & 1.1 & $\mathbf{1}$ & 1.3 & & \\
\noalign{\smallskip}\hline
\end{tabular}
\end{table}

\begin{figure}
    \includegraphics[width=4in]{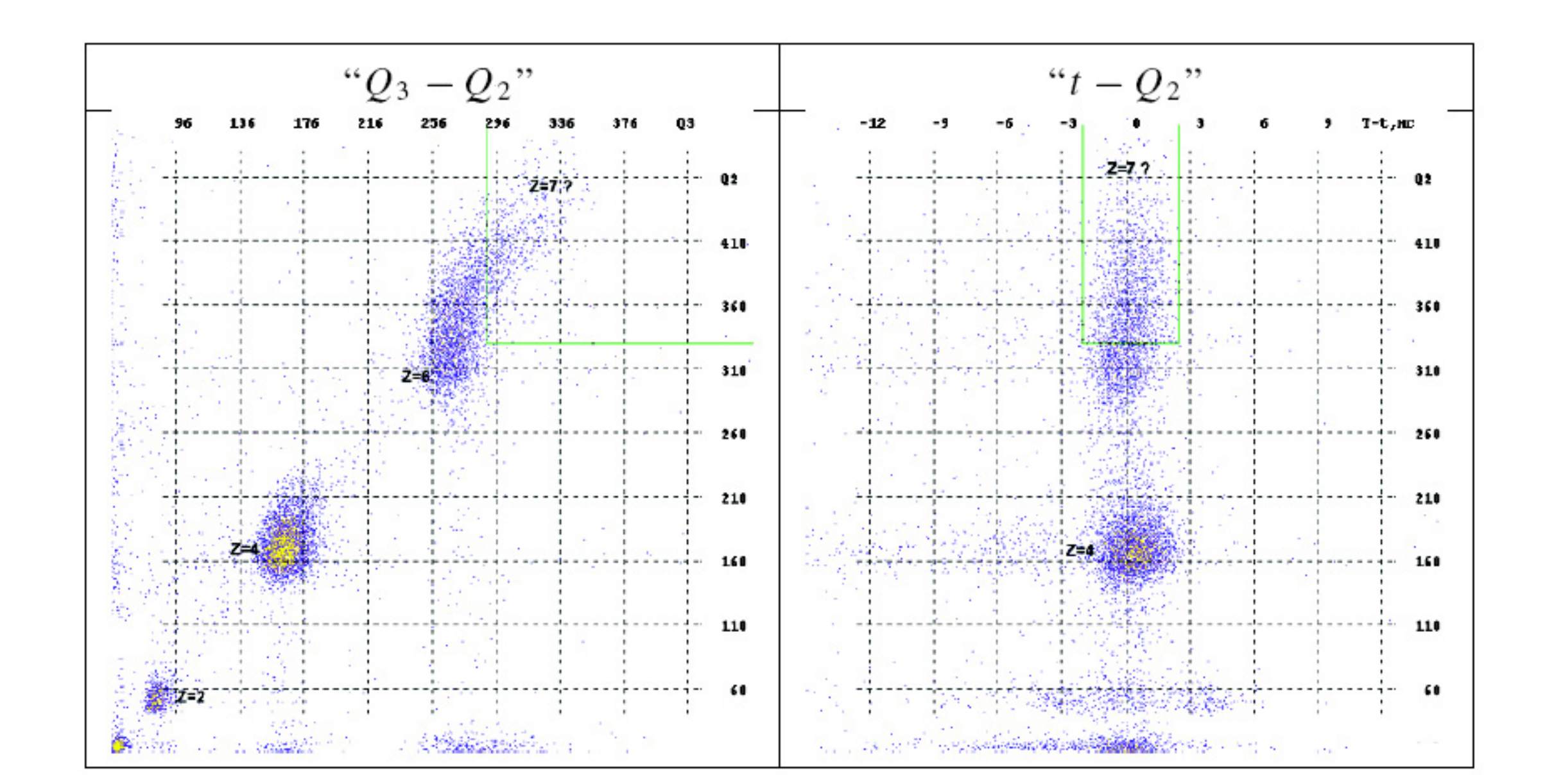}
    \caption{\label{Fig:7} Correlation diagrams of events registered in a secondary beam $^{12}$C$+$A$\rightarrow$(Z/A$=7/12)+...$ at p$=$2.0A~GeV/$c$. Left: S$_3-$S$_2$ analyzing counter signals correlation (QDC channels). Right: time-of-flight flight (ns, faster fragments right) and the S$_2$ signal correlation.}
\end{figure}

\begin{figure}
    \includegraphics[width=5in]{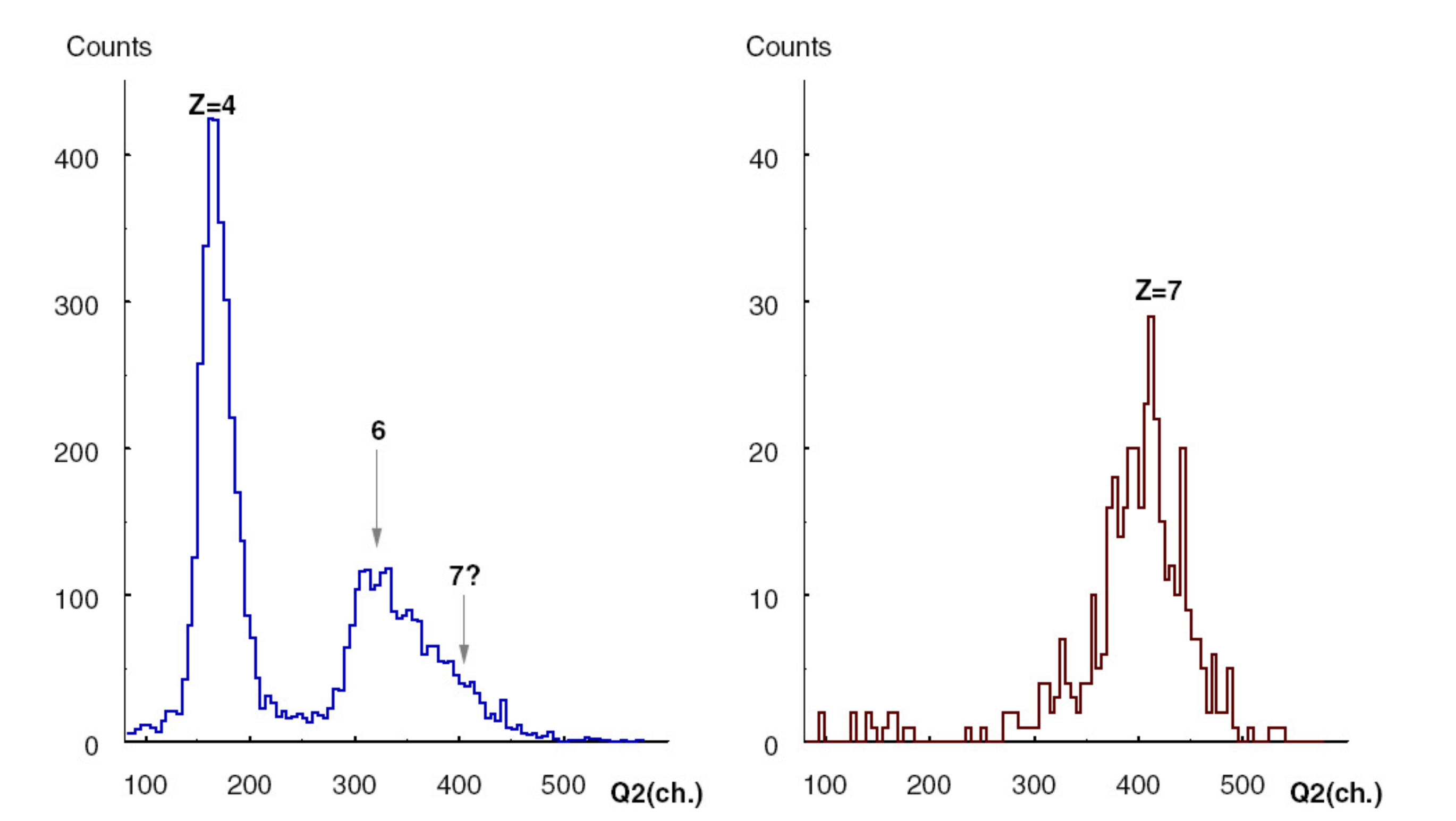}
    \caption{\label{Fig:8} Signal spectra of the 5~mm S$_2$ analyzer. Left: a histogram of events with a TOF cut only vertical solid lines in the right plot of Fig.~\ref{Fig:7}. Right: with an additional low limit cut on the S$_3$ signal.}
\end{figure}

\indent \par

\indent The work was supported in part by the Russian Foundation for Basic Research grant 04-02-17151 and grants from the JINR Plenipotentiaries of Czech Republic (2006), Slovak Republic (2004) and Romania (2002).\par

\end{document}